# Characterization of the ATLAS Liquid Argon Front-End ASIC ALFE2 for the HL-LHC upgrade

D. Matakias,[a,1] G. Carini,[b] H. Chen,[a] M. Dabrowski,[c,d] G. Deptuch,[b] L. Duflot,[e] J. Kierstead,[b] T. Liu,[a] H. Ma,[a] N. Morange,[e] S. Rescia,[b] S. Tang,[a] and H. Xu[a]

[a] *Physics Department, Brookhaven National Laboratory,*
  *Upton, NY 11973, United States of America*

[b] *Instrumentation Division, Brookhaven National Laboratory,*
  *Upton, NY 11973, United States of America*

[c] *Formerly at Instrumentation Division, Brookhaven National Laboratory,*
  *Upton, NY 11973, United States of America*

[d] *KU Leuven,*
  *Leuven, 3000, Belgium*

[e] *IJCLab, Université Paris-Saclay,*
  *CNRS/IN2P3, Orsay, 91405, France*
  *E-mail*: dimitrios.matakias@bnl.gov

ABSTRACT: ALFE2 is an ATLAS Liquid Argon Calorimeter (LAr) Front-End ASIC designed for the HL-LHC upgrade. ALFE2 comprises four channels of pre-amplifiers and CR-(RC)$^2$ shapers with adjustable input impedance. ALFE2 features two separate gain outputs to provide 16-bit dynamic-range coverage and an optimum resolution. ALFE2 is characterized using a Front-End Test Board (FETB) based on a Zynq UltraScale+ MPSoC and two octal-channel 16-bit high-speed ADCs. The test results indicate that ALFE2 fulfills or greatly exceeds all specifications on gain, noise, linearity, uniformity, and radiation tolerance.

---

[1] Corresponding author.

**Contents**

## 1. Introduction

The High-Luminosity Large Hadron Collider (HL-LHC) upgrade, scheduled to be completed for operation in 2029, is expected to increase the instantaneous luminosity up to $7.5 \times 10^{34}$ cm$^{-2}$s$^{-1}$. During the upgrade, the readout electronics of the ATLAS Liquid Argon calorimeter (LAr) will be replaced for higher data rates and increased radiation tolerance [1]. An ATLAS LAr Front-End (ALFE), a large-dynamic-range, low-noise, high linearity Pre-Amplifier (PA) and shaper ASIC, has been developed for the HL-LHC upgrade [2]. ALFE2 is the final and production version of ALFE. A small batch of ALFE2 chips has been characterized with performance exceeding specifications by a large margin. A total of about 80,000 ALFE2 chips will be manufactured for the final production. The characterization results, including the performance in lab conditions and radiation tolerance, are presented in this paper.

## 2. Design of ALFE2

A block diagram of ALFE2 is shown in Figure **1(a)**. ALFE2 integrates quad-channel transimpedance PAs and CR-(RC)$^2$ shapers. A key feature of ALFE2 is to cover a dynamic range of 16 bits. To achieve such a high dynamic range, each channel of ALFE2 has two outputs with different gains, Low Gain (LG) and High Gain (HG). Both outputs can be simultaneously read out. Each HG or LG output has a CR-RC shaping stage and an RC shaping stage. To be compatible with the current trigger system [3], ALFE2 includes a Trigger-Sum (TS) output based on the low gain outputs of the PAs. A CR-RC shaping is applied in the TS output since further anti-aliasing is performed in the trigger chain. The ATLAS LAr calorimeter readout uses a transmission line coupled PA at room temperature. The front-end PA, therefore, must have a controlled input impedance to terminate the line. The nominal detector capacitance in the high precision region is 330 pF at 50 Ω (front section) and 1.5 nF at 25 Ω (middle and back sections), respectively. Detailed specifications [4] are summarized in Table 1 of Section 4. Channels can individually be powered off.

ALFE2 is implemented as a fully differential circuit architecture. The differential architecture doubles the voltage swing, thus making the ASIC much less sensitive to common-mode noise. The PA, which performs the single-ended to differential conversion plays a crucial role in achieving the required noise and linearity. Each PA consists of an input stage (A0) and an output stage (A1). The controlled terminating input resistance is realized by means of an asymmetric negative feedback [5] as shown in Figure **1(a)**. To achieve low noise while minimizing power, the input stage utilizes an inverter-based differential circuit [6] as shown in Figure **1(b)**. The negative feedback paths utilize only passive components, making it possible to implement linear response and stable termination. ALFE2 operates with two power domains, 1.2 V for the input pair of the PAs and the shapers and 2.5 V for the output stage of the PAs. This dual-power-domain design not only minimizes power consumption but also enhances the transimpedance gain.

The gain of the TS output, the timing constant of shapers, and the output baseline voltage are configurable. ALFE2 has an I$^2$C interface with 8-bit internal registers for configuration. The I$^2$C interface operates in conjunction with a 40 MHz system clock. The I$^2$C interface employs a Triple Modular Redundancy (TMR) to enhance resilience against Single-Event-Effects (SEEs). The registers support an auto-refresh feature. With this feature, an error bit that does not agree with the other two bits in a triplicated cell is set to the value of the other two bits.

The ALFE2 ASIC is manufactured in a 130 nm CMOS process. The silicon die measures 4.55 mm in width and 5.00 mm in length. The chip is housed in a Ball Grid Array (BGA) package featuring 196 pins at 0.8 mm pitch. The package dimensions are 12 mm × 12 mm.

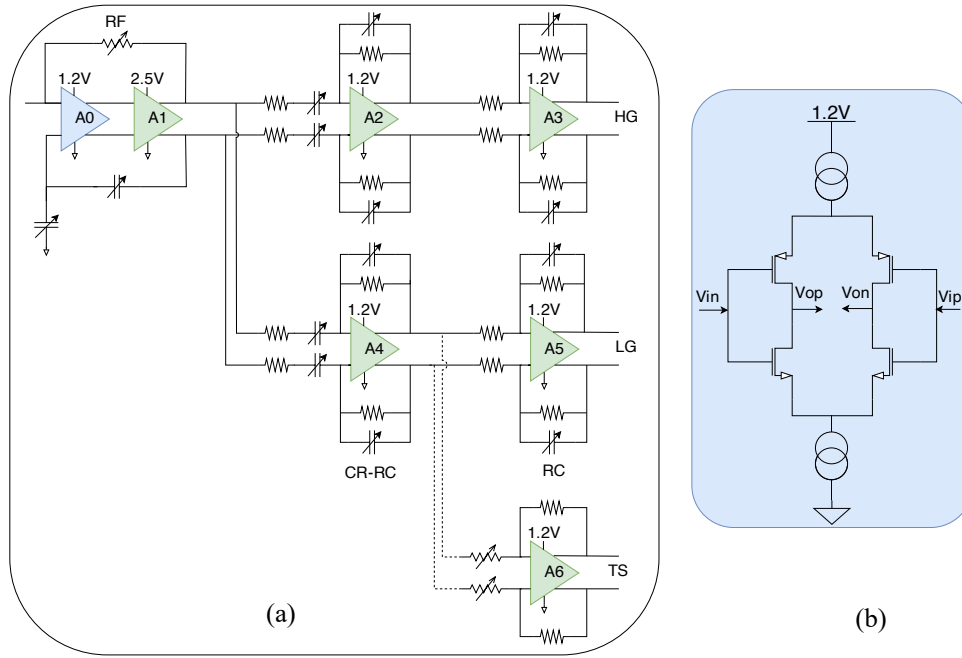

Figure 1. Block diagram (a) of ALFE2 and schematic (b) of the input stage A0.

## 3. Test setup of ALFE2

The block diagram of the test setup is shown in Figure 2(a). The ALFE2 chip under test is soldered or held in a socket mounted on a test board. The ALFE2 test board is a daughterboard of a Front-End Test Board (FETB) [7] with an FPGA Mezzanine Card (FMC) connector. The daughterboard can be directly connected to the FETB. An FMC extension cable can also be inserted between the ALFE2 test board and the FETB for irradiation tests.

A calibration pulser board generates a precise input calibration signal for ALFE2. The calibration pulser board consists of a 16-bit Digital-to-Analog Converter (DAC) (Texas Instruments DAC8830) and a low-offset Operational Amplifier (OA) ASIC. The OA ASIC, designed by IN2P3/OMEGA, is employed on the existing ATLAS LAr calibration board [8]. The FETB uses a Serial Peripheral Interface (SPI) to configure the DAC on the Pulser board. The FETB also sends a command pulse on the calibration pulser to output the calibration signal.

The calibration signal is sent to the ALFE2 test board through a calibration distribution board that consists of a network of capacitors and resistors to mimic the LAr calorimeter load. The calibration board also fans out a single calibration pulse into eight outputs. It is connected to the ALFE2 test board using 1.5 m SMA cables to approximate the connections to the detector. The impedance of each cable is 50 Ω. Two cables are used in parallel with T connectors to implement a 25 Ω impedance.

The FETB is a versatile data acquisition system that integrates analog circuits and an FPGA. The FETB features dual octal-channel 16-bit ADCs (Texas Instruments ADS52J65) with an Effective Number of Bits (ENOB) of 13 and a sampling rate of up to 125 MS/s (40 MS/s is used in the test). The FETB accommodates a System-on-Module (Enclustra Mercury XU1) powered by a Xilinx Zynq UltraScale+™ Multi-Processor System on Chip (MPSoC). The FETB has a lightweight PetaLinux operating system running on the embedded ARM processor and enables real-time analysis and data streaming through an Ethernet interface.

The test setup is used in characterization measurements, irradiation tests, and Quality Control (QC) evaluations. A photo of the test setup is shown in Figure **2(b)**.

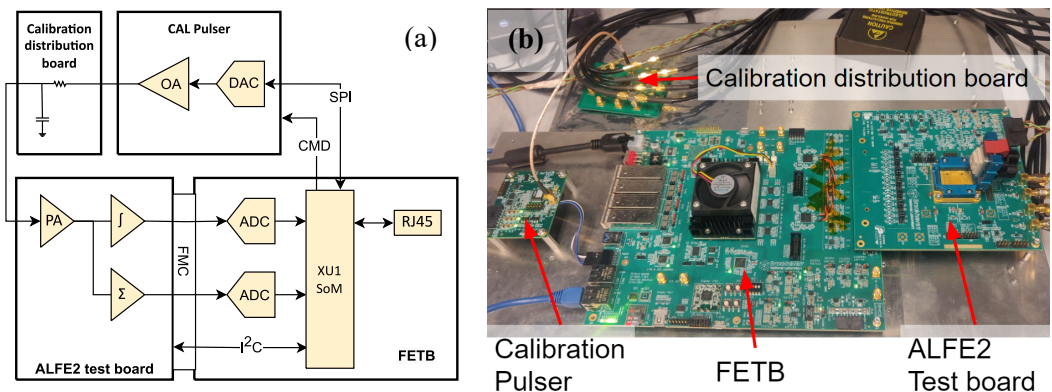

Figure 2. Block diagram (a) and photo (b) of the ALFE2 test setup.

## 4. Test results of ALFE2

A small batch of ALFE2 chips has been fabricated and packaged. A series of tests have been performed on chips from a single wafer to ensure conformity with all specifications and assess uniformity and yield.

The transient responses of the LG outputs at 25 Ω and various input currents are shown in Figure **3(a)**. The ALFE2 outputs were captured with the ADCs at 40 MS/s, the same sampling frequency used in the ATLAS LAr calorimeter. In the figure, each curve is reconstructed from 18 overlapped waveforms at different phases. The gain was extracted with a polynomial fit centered around the pulse peak, while the peaking time was calculated from 5% to 100% of the peak. The relative gain ratio between the HG and the LG outputs was measured at 22.9 on average. The gain of the TS output relative to the LG ones is programmable to be 1, 2, or 3 (±3%). The peaking time is programmable from 29 ns to 46 ns for HG/LG outputs at 50 Ω, and from 37 ns to 57 ns for the HG/LG outputs at 25 Ω. The trigger sum peaking time is configurable from 26 to 34 ns at 50 Ω, and from 34 to 45 ns at 25 Ω, respectively.

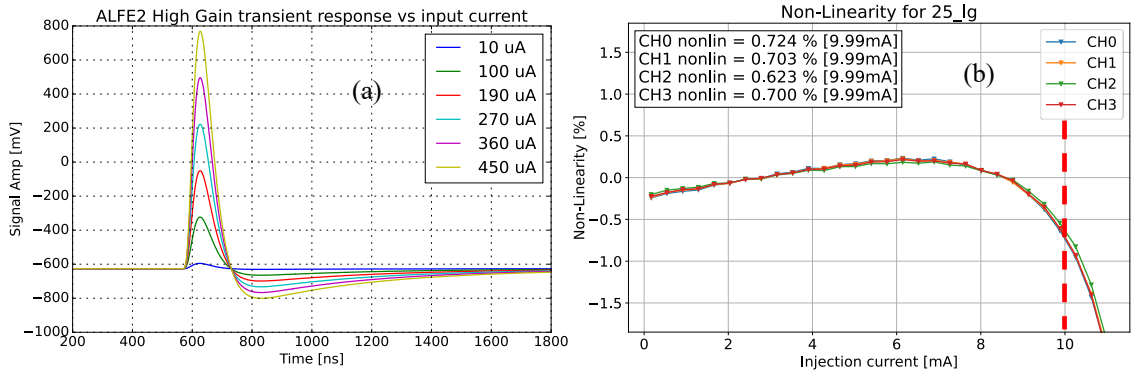

Figure 3. Waveforms (a) and INL (b) of the LG outputs (b) at 25 Ω.

The Integral Non-Linearity (INL) measurement of ALFE2 was performed. A typical measurement of the INL for the LG outputs at 25 Ω is shown in Figure **3(b)**. The INL of the LG outputs is below ±0.62% across the full dynamic range and ±0.21% within 80% of the dynamic range. The INL of the HG outputs is less than ±0.13%. The INLs of the TS output are below ±0.41%. All measured INLs are less than half of the specifications.

The calculation of the Equivalent Noise Current (ENI) in ALFE2 involves dividing the RMS noise derived from the pedestal of the chosen output by the gain. A typical relationship between the ENI and the peaking time at the HG outputs at 25 Ω is depicted in Figure **4(a)**. The ADC noise contribution was quadratically subtracted to generate these measured ENIs. The ENIs were measured to be 48 nA and 153 nA with peaking times of 38 ns at 50 Ω and 46 ns at 25 Ω. The ENI histogram of 120 pre-production chips at the HG outputs at 25 Ω is shown in Figure **4(b)**. The ENI values shown in Figure 4(b) are slightly lower than those shown in Figure 4(a) due to the inclusion of ADC anti-alias filters. One chip with an ENI of 170 nA, slightly higher than others, is not shown in Figure **4(b)** because of out of scale.

The power dissipation is 610 mW on average or 153 mW per channel. The power dissipation of all chips is within the specification of 650 mW.

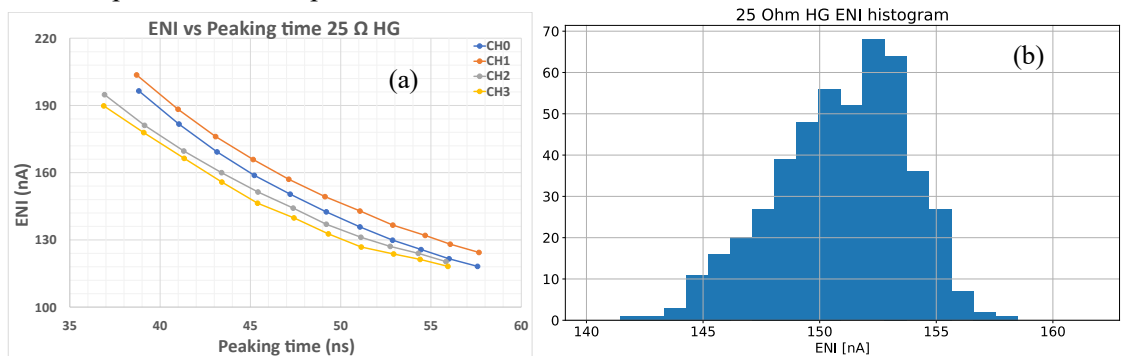

Figure 4. ENIs of a typical chip (a) and ENI histogram of 120 chips (b) of the HG outputs at 25 Ω.

The ALFE2 underwent radiation hardness testing (TID) at Brookhaven National Laboratory, utilizing a $^{60}$Co gamma-ray source at a dose rate of 0.28 Gy/s. In this evaluation, two separate samples were subjected to doses of 7 kGy and 15 kGy, respectively. The chip performance, temperature, input voltages, and internal register values were monitored in real-time. Less than 1% performance change was measured on the gain and the TS baseline after the dose was beyond the specification of 1.4 kGy, whereas no change was observed on other metrics.

A SEE test, using a 226 MeV proton beam was performed at Massachusetts General Hospital in Boston. During the test, almost all (15 out of 16 bytes) internal registers except the one byte that enabled/disabled channels were continuously written with random numbers and read back. Four chips were irradiated to a total fluence of $1.7 \times 10^{14}$ cm$^{-2}$, 17 times of the specification, with the auto-refresh feature enabled, and no Single-Event-Upsets (SEUs) were detected in any chip. The results of the four chips are combined, and the SEU cross-section is estimated to be less than $1.4 \times 10^{-16}$ cm$^2$ at the 95% confidence level. The error rate for the entire ATLAS LAr calorimeter is extrapolated to be under 5 errors per day (95% confidence level).

The measurement values in Table 1 are averaged over 120 ASICs, or a total of 480 channels. All measurements meet and significantly exceed the specifications. Five chips were rejected. The rejection criteria were established using the mean and sigma values for each metric as a reference.

**Table 1**. Specifications and measurements of ALFE2.

| Parameters | | Specifications | Measurements | |
|---|---|---|---|---|
| Maximum input current (mA) | | HG: 0.08 (50 Ω), 0.45 (25 Ω) LG: 2 (50 Ω), 10 (25 Ω) | | |
| Nominal capacitive load (pF) | | 330 pF (50 Ω), 1500 pF (25 Ω) | | |
| Relative gain | HG/LG | 22±5 | 22.9 | |
| | TS ×3/×1 | 1 or 3 (±5%) | 1, 2, or 3 (±3%) | |
| Peaking time (5%-100%, ns) | HG/LG 50 Ω | 38±5 | programmable within 29-46 | |
| | HG/LG 25 Ω | 46±5 | programmable within 37-57 | |
| | TS 50 Ω | 27±5 | programmable within 26-34 | |
| | TS 25 Ω | 36±5 | programmable within 34-45 | |
| INL | HG | <0.2% | 0.13% | |
| | LG 80% range | <0.5% | 0.21% | |
| | LG full range | <5% | 0.62% | |
| | TS | <2% | 0.41% | |
| ENI (nA)* | HG 50 Ω | <120 | 48 | 50 |
| | HG 25 Ω | <350 | 153 | 168 |
| Power dissipation (mW) | | <650 | 610 | |

*The ADC noise contribution has been quadratically subtracted on the left column of the measurements. The right column includes the ADC contribution.

## 5. Conclusion

The ALFE2 ASIC, designed for the ATLAS LAr calorimeter upgrade, has been extensively characterized. The measured INLs of the HG outputs are 0.09%. The INLs of the LG outputs are 0.21% in the 80% dynamic range and less than 0.62% in the full dynamic range. The measured

ENIs are 48 nA and 153 nA at 50 Ω and 25 Ω, respectively. The power dissipation is 610 mW. So far, 120 pre-production chips have been tested and all meet the specifications. The test results indicate that ALFE2 fulfills or greatly exceeds all specifications on gain, noise, linearity, uniformity, and radiation tolerance. Based on the characterization results, about 80,000 ALFE2 chips will be produced for the HL-LHC upgrade.

**Acknowledgments**

We wish to extend our sincere appreciation to the individuals who contributed to the success of the ALFE2 chip design project. First, we would like to acknowledge G. de Geronimo who designed HLC1, a predecessor to the ALFE. We would like to express our gratitude to the IN2P3/OMEGA team, comprising Sylvie Blin, Selma Conforti, Christophe de la Taille, Mowafak El Berni, and Nathalie Seguin-Moreau. In addition, we acknowledge the contributions made by Vamshi Manthena and Sandeep Miryala from the Instrumentation Division at Brookhaven National Laboratory and Stefano Michelis from CERN. Their collaboration played a vital role in completing the ALFE2 design. Finally, we would like to thank Ethan Cascio from the Proton Therapy Center at Massachusetts General Hospital in Boston.